\documentclass[fleqn,twoside]{article}
\usepackage{espcrc2}
\usepackage{epsfig}
\usepackage{latexsym}

\title{{Lattice QCD at finite isospin density and/or temperature}\thanks{Talk
presented by D.~K.~Sinclair, Lattice2003, Tsukuba.}}

\author{J.~B.~Kogut\address{Department of Physics, University of Illinois,
1110 West Green Street, Urbana, IL 61801, USA}\thanks{Supported in part by NSF
grant NSF PHY-0102409.} and
D.~K.~Sinclair, \address{HEP Division, Argonne National Laboratory, 9700 South
Cass Avenue, Argonne, IL 60439, USA}\thanks{Supported by US Department of 
Energy, High Energy Physics Division, under contract W-31-109-ENG-38.}}

\begin{document}

\begin{abstract}
We simulate two-flavour lattice QCD with at a finite chemical potential $\mu_I$
for isospin, and finite temperature. At small $\mu_I$, we determine the position
of the crossover from hadronic matter to a quark-gluon plasma as a function of
$\mu_I$. At larger $\mu_I$ we observe the phase transition from the superfluid 
pion-condensed phase to a quark-gluon plasma, noting its change from second
order to first order as $\mu_I$ is increased. We also simulate two-flavour 
lattice QCD at zero quark mass, using an action which includes an additional
4-fermion interaction, at temperatures close to the chiral transition on $N_t=8$
lattices.
\end{abstract}

\maketitle

\section{Introduction}

QCD at finite temperature and/or densities is relevant to the physics of the
early universe, neutron stars and relativistic heavy-ion collisions --- RHIC
and the CERN heavy-ion program.

Because of the difficulties involved with simulating QCD at finite chemical
potential $\mu$ for quark number with its complex fermion determinant, we are
simulating $N_f=2$ lattice QCD with a finite chemical potential $\mu_I$ for
isospin ($I_3$), which has a real positive fermion determinant \cite{ks1}. We
are currently studying the $\mu_I$ dependence of the finite temperature
transition \cite{ks2}. At small $\mu$, the Bielefeld-Swansea collaboration has
observed that the phase of the fermion determinant is sufficiently well
behaved that the $\mu$ and $\mu_I$ dependence of the transition are identical
\cite{bs1}. Our predictions are in good agreement with those of de~Forcrand
and Philipsen \cite{dfp1}.

At large values of $\mu_I$ ($\mu_I > m_\pi$), the low temperature phase is
characterized by a pion condensate. The finite temperature transition is now a
true phase transition, which appears to be second order for lower $\mu_I$
values and first order at higher $\mu_I$ values.

We also simulate lattice QCD with an irrelevant chiral 4-fermion term which
allows us to simulate at zero quark mass, giving us direct access to the 
critical exponents at the finite temperature transition \cite{xqcd}.
We are using $16^3 \times 8$ and $24^3 \times 8$ lattices for these
simulations.

In section~2 we present preliminary results of our finite $\mu_I$ and
temperature simulations. Section~3 gives some preliminary graphs from our 
$N_t=8$ finite-temperature simulations using our modified action. Finally we
present our conclusions and indicate further avenues of research in section~4.

\section{QCD at finite $\mu_I$ and temperature}

The quark part of our lattice action is
\begin{equation}
S_f=\sum_{sites} \left[\bar{\chi}[D\!\!\!\!/(\frac{1}{2}\tau_3\mu_I)+m]\chi
                   + i\lambda\epsilon\bar{\chi}\tau_2\chi\right].
\end{equation}
For simulations at $\mu_I < m_\pi$, we set the symmetry breaking parameter
$\lambda=0$.

\begin{figure}[htb]
\epsfxsize=3in
\vspace{-0.2in}
\centerline{\epsffile{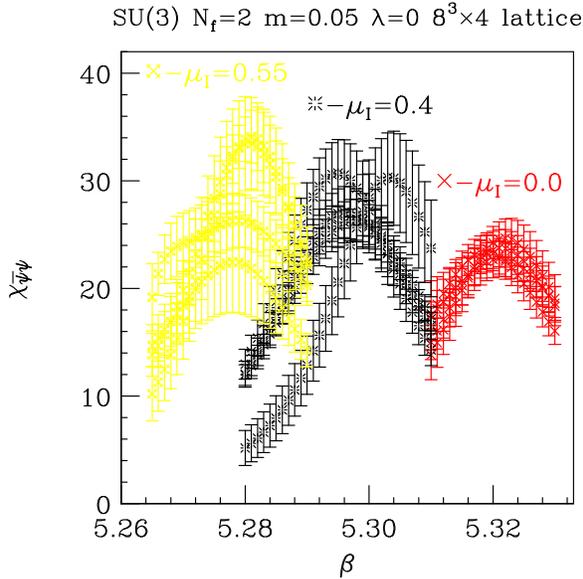}}
\vspace{-0.2in}
\caption{Ferrenberg-Swendsen reweighted chiral susceptibilities for 
$\mu_I=0.0,\: 0.4, \: 0.55$.}
\label{fig:fs}
\end{figure}

We perform simulations on an $8^3 \times 4$ lattice with $m=0.05$, $\lambda=0$
and $0 \le \mu_I \le 0.55$, for a set of $\beta$s covering the crossover region
for each $\mu_I$. We measure the chiral condensate, the plaquette, the Wilson
Line and the isospin density and their corresponding susceptibilities for each
set of parameters. The position of the crossover was obtained as the peak of
the susceptibilities. This was determined using Ferrenberg-Swendsen reweighting
techniques \cite{fs}. Figure~\ref{fig:fs} shows the chiral susceptibilities
obtained from such reweightings for 3 of the 7 $\mu_I$ values we use. The
multiple values for each $\mu_I$ are the results of using distributions from
several $\beta$s close to the peak.

\begin{figure}[htb]
\epsfxsize=3in
\centerline{\epsffile{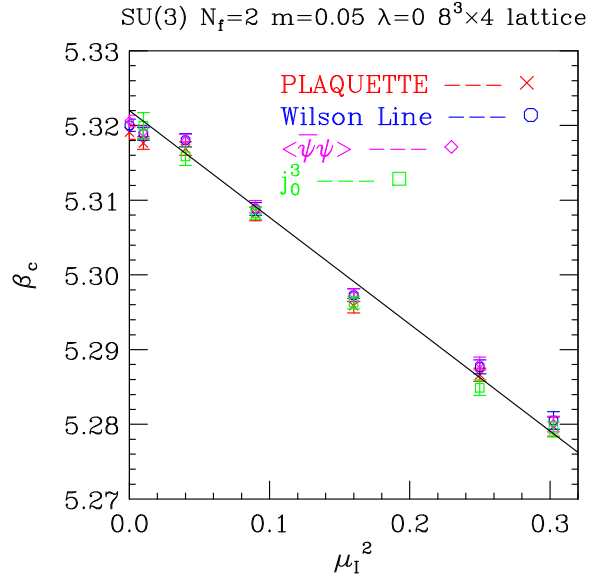}}
\vspace{-0.2in}
\caption{Position of the crossover as functions of $\mu_I^2$.}
\vspace{-0.2in}
\label{fig:beta_c}
\end{figure}

The crossover $\beta$ values, $\beta_c$, obtained from the peaks in the
susceptibilities for each of the 4 observables, are plotted in 
figure~\ref{fig:beta_c} as functions of $\mu_I^2$, since the leading term is
expected to be quadratic in $\mu_I$. We note that the 4 predictions for each
$\mu_I$ appear consistent within errors, indicating that this is a reasonable
definition of the position of the crossover. The straight line 
\begin{equation}
\beta_c=5.322-0.143\mu_I^2
\end{equation}
in this figure is only meant as a rough guide. Using the relation
$\beta_c(\mu)=\beta_c(\mu_I=2\mu)$, which should hold for small $\mu,\:\mu_I$ 
our results are consistent with those of de~Forcrand and Philipsen \cite{dfp1}.

We are repeating these simulations for $m=0.1$ and $m=0.2$ where the critical
$\mu_I$s, $\mu_c=m_\pi$ are larger. In no case does there appear to be a 
critical end point, beyond which the transition becomes first order, for 
$\mu_I < m_\pi$.

We are extending our $m=0.05$ simulations to $\mu_I > m_\pi$. While the thermal
phase transition where the pion condensate evaporates appears to be second order
for $\mu_I=0.6$, simulations on $16^3 \times 4$ lattices show indications of
the metastability expected for a first order transition, at $\mu_I=0.8$.

\section{``$\chi$QCD'' at finite temperature}

We simulate lattice QCD with 2 flavours of massless staggered quarks and an
irrelevant chiral 4-fermion interaction which allows us to run at zero quark
mass, at finite temperature on $16^3 \times 8$ and $24^3 \times 8$ lattices.
Using $N_t=8$ should free us from the lattice artifacts which were present
for $N_t=4,6$ \cite{xqcd}.

Having massless quarks should enable us to extract the critical exponent
$\beta_m$ which describes the vanishing of the chiral condensate as the
chiral transition is approached from below, and the critical $\beta$ 
($\beta_c$). Running at $\beta_c$ with small quark masses will give us the
critical exponent $\delta$.

\begin{figure}[htb]
\epsfxsize=3in
\vspace{-0.2in}
\centerline{\epsffile{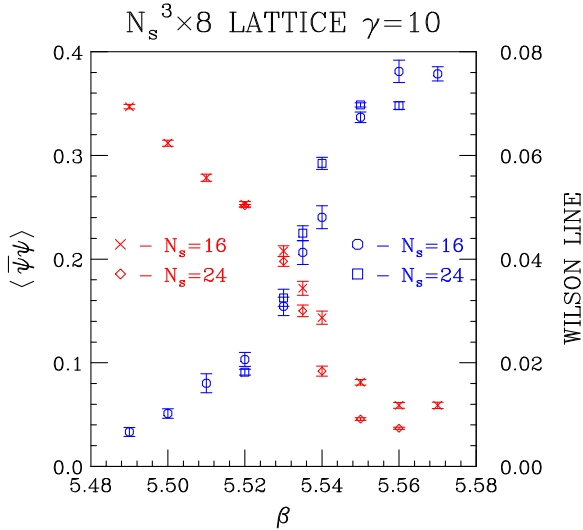}}
\vspace{-0.2in}
\caption{Chiral condensate and Wilson line of ``$\chi$QCD'' as functions of
$\beta$ on $N_t=8$ lattices.}
\label{fig:XQCD}
\end{figure}

Preliminary results for the chiral condensate and Wilson Line are given in
figure~\ref{fig:XQCD}. These clearly indicate that the transition occurs in the
range $5.530 < \beta_c < 5.545$ and appears sufficiently smooth to be second 
order.

\section{Conclusions}

We have determined the $\mu_I$ dependence of the finite temperature transition
for 2-flavour lattice QCD. For small $\mu$, the fluctuations of the phase of
the fermion determinant are small enough that the $\mu$ and $\mu_I$ dependence
of this transition are the same \cite{bs1}. The $\mu_I$ dependence we measure
predicts a $\mu$ dependence consistent with that obtained from simulations
with imaginary $\mu$ \cite{dfp1}. We will check if it is also consistent with
series expansions around $\mu=0$ \cite{bs1}. Our simulations at 3 different
quark masses indicate that the transition is a crossover for $\mu_I < m_\pi$.
At higher $\mu_I$, where the the pion condensate evaporates at the transition,
we see evidence for a change to first order behaviour, but it is unclear if
this is related to the critical endpoint expected at finite $\mu$.

Others have estimated the position of the critical endpoint at finite $\mu$
for the more physical $2+1$-flavour QCD \cite{fk}, the closely related
3-flavour QCD \cite{bs2,dfp2} and 4-flavour QCD \cite{mpl}. We are extending
our simulations to 3-flavours, where one can tune the quark mass to make
the critical endpoint as close to $\mu=0$ as desired. If the endpoint is at
small enough $\mu$, we expect a critical endpoint at the corresponding
$\mu_I$, giving another estimate for the position of this endpoint.

We are using the ``$\chi$QCD'' action to determine the critical exponents for
the chiral transition of finite temperature QCD, on $N_t=8$ lattices.

We thank F.~Karsch, P.~de Forcrand, O.~Philipsen and D.~Toublan for helpful
discussions. These simulations were performed on the IBM SP at NERSC and the
Jazz cluster at Argonne.


\begin{thebibliography}{9}
\bibitem{ks1}
J.~B.~Kogut and D.~K.~Sinclair, Phys.\ Rev.\ D {\bf 66} (2002) 034505.
\bibitem{ks2}
J.~B.~Kogut and D.~K.~Sinclair, Nucl.\ Phys.\ B\ Proc.\ Suppl.\ {\bf 119} 
(2003) 556;
J.~B.~Kogut and D.~K.~Sinclair, talk presented by D.~K.~Sinclair at ``Finite
Density QCD at Nara'' (2003).
\bibitem{bs1}
C.~R.~Allton {\it et al.}, Phys.\ Rev.\ D {\bf 66} (2002) 074507;
Phys.\ Rev.\ D {\bf 68} (2003) 014507.
\bibitem{dfp1}
P.~de Forcrand and O.~Philipsen, Nucl.\ Phys.\ B {\bf 642} (2002) 290.
\bibitem{xqcd}
J.~B.~Kogut, J.~F.~Lagae and D.~K.~Sinclair, Phys.\ Rev.\ D {\bf 58} (1998) 
034504;
J.~B.~Kogut and D.~K.~Sinclair, Phys.\ Lett.\ B {\bf 492} (2000) 228;
Phys.\ Rev.\ D {\bf 64} (2001) 034508;
arXiv:hep-lat/0211008 (2002).
\bibitem{fs}
A.~M.~Ferrenberg and R.~H.~Swendsen, Phys.\ Rev.\ Lett. {\bf 63} (1989) 1195.
\bibitem{fk}
Z.~Fodor and S.~D.~Katz, Phys.\ Lett.\ B {\bf 534} (2002) 87;
\ JHEP {\bf 0203} (2002) 014;
Z.~Fodor, S.~D.~Katz and K.~K.~Szabo, Phys.\ Lett.\ B {\bf 568} (2003) 73.
\bibitem{bs2}
C.~R.~Allton {\it et al.}, poster presented by F.~Karsch at ``Lattice2003''
(2003).
\bibitem{dfp2}
P.~de Forcrand and O.~Philipsen, arXiv:hep-lat/0307020 (2003).
\bibitem{mpl}
M.~D'Elia and M.~P.~Lombardo, Phys.\ Rev.\ D {\bf 67} (2003) 014505.
\end{thebibliography}
\end{document}